# Measuring Social Media Activity of Scientific Literature: An Exhaustive Comparison of Scopus and Novel Altmetrics Big Data

Saeed-Ul Hassan[a], Mubashir Imran[a], Uzair Gillani[a], Naif Radi Aljohani[b], Timothy D. Bowman[c], Fereshteh Didegah[d]

[a] Information Technology University, 346-B, Ferozepur Road, Lahore, Pakistan
E-mail address: saeed-ul-hassan@itu.edu.pk, Tel: + 92-322-228-9756

[b] Faculty of Computing and Information Technology, King Abdulaziz University, Jeddah, Kingdom of Saudi Arabia
E-mail address: nraljohani@kau.edu.sa

[c] School of Library and Information Science, Wayne State University, Detroit, MI, United States
E-mail address: timothy.d.bowman@wayne.edu

[d] Faculty of Communication, Art and Technology, Simon Fraser University, Vancouver, BC, Canada
E-mail address: fdidegah@sfu.ca

**Abstract**

This paper measures social media activity of 15 broad scientific disciplines indexed in Scopus database using Altmetric.com data. First, the presence of Altmetric.com data in Scopus database is investigated, overall and across disciplines. Second, the correlation between the bibliometric and altmetric indices is examined using Spearman correlation. Third, a zero-truncated negative binomial model is used to determine the association of various factors with increasing or decreasing citations. Lastly, the effectiveness of altmetric indices to identify publications with high citation impact is comprehensively evaluated by deploying Area Under the Curve (AUC) - an application of receiver operating characteristic. Results indicate a rapid increase in the presence of Altmetric.com data in Scopus database from 10.19% in 2011 to 20.46% in 2015. A zero-truncated negative binomial model is implemented to measure the extent to which different bibliometric and altmetric factors contribute to citation counts. Blog count appears to be the most important factor increasing the number of citations by 38.6% in the field of Health Professions and Nursing, followed by Twitter count increasing the number of citations by 8% in the field of Physics and Astronomy. Interestingly, both Blog count and Twitter count always show positive increase in the number of citations across all fields. While there was a positive weak correlation between bibliometric and altmetric indices, the results show that altmetric indices can be a good indicator to discriminate highly cited publications, with an encouragingly AUC= 0.725 between highly cited publications and total altmetric count. Overall, findings suggest that altmetrics could better distinguish highly cited publications.

# Introduction

Online social media applications have attracted a tremendous number of users by providing them with a unique context in which to interact with like-minded people (Priem and Hemminger 2010; Wouters and Costas 2012). Social media networks afford users the ability to share ideas and receive an immediate response to their sharing activities. Due to its rapid response capabilities, social media applications have attracted the attention of the scientific community, who, parallel to traditional (i.e. bibliometrics) forms of scholarly communication, are now using these online contexts to disseminate research in their daily scholarly practices (Thelwall et al. 2013).

In 2010, the term altmetrics was proposed as another form of collecting article level metrics in a manner that would allow for more timely measurements of interest in scholarly documents and as a means to filter the vast amount of information being disseminated online (Priem et al. 2010). Since the introduction of the term altmetrics to the Scientometrics community, scholars have been exploring the possible analytics that this online activity can offer and the impact it may have on the diverse communities within and outside the academic community including clinicians, practitioners, and the general public (see Sugimoto et al. (2017) for an extended review of altmetrics literature).

To be specific, altmetric data is used to track the use of scientific research in a variety of online platforms including, but not limited to, news sites, social media platforms, blogs, video sites, and reference management tools. In this way, altmetrics analyzes the real-time sharing of scientific documents based on various online actions, which can include comments, discussions, likes, shares, and bookmarks. (Zahedi et al. 2014).

Social media applications have noticeably impacted scholarly communication behaviors and expectations. Scholars may discuss and share their work on Twitter using a hashtag to signal that their work is relevant for a specific audience. Similarly, Facebook and Google+ can be utilized to share scholarly information within and outside of a user's immediate social network. Scholars are using social reference managers, such as Mendeley or Zotero, to organize academic references and share document metadata and tags. In addition, CiteULike and Pinterest has been used to bookmark, or pin, scholarly documents related to a user's discipline or interests (Haustein and Siebenlist 2011; Nielsen 2007). Among the altmetric platforms being analyzed by scholars, Mendeley has been shown to be of significant importance. Zahedi, et al. (2013) found that approximately 63% percent of overall metrics were linked with Mendeley readership, while other altmetric sources demonstrate a very marginal link. Furthermore, a very moderate correlation ($r = 0.49$) was detected between Mendeley readership count and citation indicators.

According to Priem et al. (2010), scholars strive to stay abreast of the most current research in their field. When using traditional citation measures, it may take several years to identify the most relevant research (as citations take time to accrue). The activity captured by altmetric researchers and data providers, however, allows for the analysis of acts in a more immediate manner (Brody et al. 2006). Online acts allow one to filter out the more popular (as indicated by online activity) research as it is published. Altmetrics seems to measure a different type of research activity from citations and is gaining interest in various fields. Although research evaluation continues to be focused on citation-based metrics, the limitations of the technique are evident (Zahedi et al. 2014; Hassan and Gilani 2016). Due to the limitations of citation and altmetrics evaluation techniques alone, a "multi-metrics approach" has been outlined that would analyze the research impact in broader aspects (Zahedi et al. 2013). According to Costas et al. (2015), altmetrics scores can be utilized in identifying highly cited publications demonstrating higher-level accuracy as compared to journal citation scores, although the level of recall is very low.

In recent years, a number of studies have been conducted to analyze the relation between altmetrics and traditional citation based indices (see e.g. Sugimoto et al. 2008; Priem et al. 2012; Bar-Ilan et al. 2012; Thelwall et al. 2013; Sud and Thelwall 2014; Haustein et al. 2014a; Haustein et al. 2014b; Yu 2017). Much of the previous work demonstrates a weak-to-high correlation between the various metrics, depending upon their dataset coverage, and authors have suggested that the researchers should conduct larger-scale studies to merge quantitative and qualitative approaches. One of the most comprehensive studies to date, by Costas et al. (2015), covers 75,569 publications indexed in the Web of Science (WoS), which supplemented the WoS data with metrics provided by Altmetric.com (http://www.altmetric.com/). Altmetric.com is a commercial tool that collects altmetric-related indices around scientific publications from online platforms including Twitter, blogs, Google+, Facebook, and various news outlets (Adie and Roe, 2013).

In this paper, a full-scale comparison of altmetric indices with selected known traditional bibliometric indices is undertaken by using more than 1.1 million publications indexed in Scopus during 2011 to 2015 and relevant altmetric information from Altmetric.com (version dateset-jun-4-2016.tar.gz). This study is similar to the works of Didegah et al. (2017), Costas et al. (2015) and Haustein et al. (2014a), which was undertaken using WoS data and a combination of WoS data and PubMed data, respectively. This work complements their findings by examining scientific literature retrieved from the Scopus database. Based on a review of the relevant literature, this work represents one of the largest studies seeking to measure the social media activity of scientific literature in relation to bibliometric indices. The objectives of this study are to investigate the following aspects:

1. What is the presence of Altmetric.com data in Scopus database overall and across the disciplines?
2. How well do altmetric indices correlate with the traditional bibliometric based indices?
3. Which, and to what extent, bibliometric and altmetric factors associate with number of citations to articles?
4. Can altmetric scores discriminate publications with a higher citation impact?
5. Which altmetric indicator represents the most significant means to discriminate highly cited publications?

**Data and methodology**

Altmetric.com shared their dataset with the authors on June 14, 2016 (version dateset-jun-4-2016.tar.gz). Each article in the dataset contains information from various streams of activities associated with it from various online platforms and can be uniquely identified through an altmetric identifier. This version of altmetric data consists of 4.5 million JSON files, with each file representing a single publication. Since Altmetric.com started data collection beginning in the second half of 2011, publications from July 2011 through June 2016 contain altmetric data. Using this data, the authors obtained 1.7 million unique publications with a reported publication date between July 2011 and December 2015 that exists both in Altmetric.com and the Scopus database.

Cross matching is performed based on DOIs (when available) and publication titles (when DOI is not available) using the Scopus API. Because the Altmetric.com data only provides altmetric web-based related indices, citation counts were collected for all 1.7 million publications using the Scopus API. Based on this cross-matching, the final dataset consisted of 1,104,275 publications that have at least one citation count and at least one captured social activity. The total number of documents published in Scopus between 2011-2015 totals 10,402,564, which was too large to analyze and model in the current study. Hence, a threshold was set in which a document had to contain at least one citation and one Altmetric.com captured event; this was necessary to apply in order to limit the Scopus dataset into a manageable collection for analyses. The citation window for each publication is from the reported time of publication to February 2017. This gives more than a year time window to the publications published in 2015. Using this final dataset, the objectives of this study can be met.

For cross disciplinary analysis, the ASJC (All Science Journals Classification) subject categories were employed; ASJC is a subject classification scheme employed by Scopus to index source titles in a structured hierarchy of disciplines and sub-disciplines. Similar to the work of Haddawy et al. (2017), the top-level 27 ASJC disciplines were merged into 15 disciplines by combining "Agricultural & Biological Sciences" and "Veterinary" into "Agricultural, Biological Sciences & Veterinary", "Business Management and Accounting", "Decision Sciences" and "Economics, Econometrics and Finance" into "Economics, Business & Decision Sciences", "Chemical Engineering", "Energy" and "Engineering" into "Engineering", "Health professions" and "Nursing" into "Health Professions & Nursing", "Immunology and Microbiology", "Neuroscience" and "Pharmacology, Toxicology and Pharmaceutics" into "Other Life & Health Sciences", and "Social Science" and "Psychology" into "Social Science". This reduced mapping allowed for normalization of the effect of source titles that are indexed across multiple top-level ASJC disciplines. For simplicity, no fractionalized counting schemes were used for publications, citations, or altmetric indicators. The "Multidisciplinary" category of ASJC was not used in this work due to the presence of source titles like Science, Nature, and PNAS.

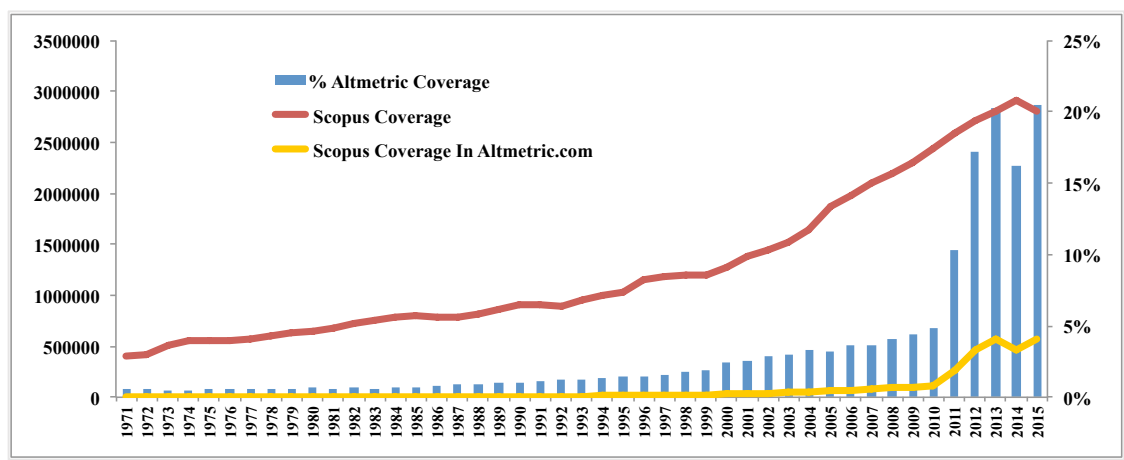

**Figure 1:** General presence of publications in Altmetric.com dataset as compared to Scopus dataset through October 31, 2015 (red line: number of articles in Scopus dataset; yellow line: number of articles present in Altmetric.com dataset whose DOI`s match.

Of the total Scopus database, 5.81% is covered by Altmetrtic.com during 1971 to 2015. Fig. 1 clearly demonstrates that the coverage of Altmetric.com's data in relation to the Scopus database has increased rapidly in recent years, particularly during 2011 to 2015.

The general presence of Altmetric.com data indexed in Scopus is 10.29% in 2011 and reaches its peak in 2015 with 20.46%. As might be expected, documents with a reported publication date between 2011-2015 have greater social activity as compared to previous years. In comparison, Haustein et al. (2014a) found that less than 10% of the 1.4 million biomedical papers indexed by PubMed and WoS were tweeted; they also found a variation by time, with papers from 2012 receiving more tweets than those from 2010.

**Table 1:** Percentage of altmetric.com data in Scopus database across disciplines, from July 2011 through December 31, 2015.

| Fields | Scopus Coverage | % Altmetric Coverage |
|---|---|---|
| Biochemistry, Genetics and Molecular Biology | 1,579,999 | 26.44 |
| Medicine and Medical Sciences | 4,253,960 | 23.55 |
| Health Professions & Nursing | 383,681 | 22.92 |
| Other Life & Health Sciences | 1,095,985 | 22.68 |
| Social Sciences | 1,432,313 | 15.55 |
| Chemistry | 1,102,103 | 11.84 |
| Environmental Science | 652,555 | 11.31 |
| Earth and Planetary Sciences | 573,728 | 9.60 |
| Economics, Business & Decision Sciences | 576,103 | 7.18 |
| Physics & Astronomy | 1,523,345 | 6.73 |
| Materials Science | 1,284,044 | 4.55 |
| Mathematics | 862,599 | 3.65 |
| Engineering | 3,362,610 | 2.94 |
| Computer Science | 1,582,557 | 2.26 |
| Agricultural, Biological Sciences & Veterinary | 13,821,961 | 1.59 |

Table 1 depicts the presence of Altmetric data across the different disciplines of Scopus, as mapped using Scopus ASJC. The field of Biochemistry, Genetics and Molecular Biology shows the most coverage by Altmetric.com at 26.44%, followed by Medicine, 26.44%, and Health Professions & Nursing, 22.92%. Agricultural, Biological Sciences & Veterinary may have the largest coverage in the Scopus data, but the presence of Altmetric.com articles categorized under Agricultural, Biological Sciences & Veterinary in Scopus are only 1.59%. The overall trend indicates that the presences of articles from the Medical and Social Sciences in Altmetric.com data found in Scopus data is greater than all other disciplines and received a greater percentage of social media attention.

**Dependent and independent variables**

The number of citations an article received is the dependent variable and the independent variables are two types of bibliometric factors and altmetric factors as follows:

- *Bibliometric factors*
    - Source Normalized Impact per Paper (*SNIP*): The SNIP[1] measures the impact of source titles by normalizing the citation potential in the field (Waltman et. al., 2013).
    - Document type: Documents retrieved from Scopus are in four different types including articles, letters, reviews, and non-citable documents. The majority of documents in all fields are article type. In order to import document type into the regression model, the types are coded as shown in Table 2.

**Table 2:** Document type and code

| Doc. Type | Type code |
|---|---|
| non-citable | 1 |
| article | 2 |
| letter | 3 |
| review | 4 |

---

[1] The SNIP 2015 data was downloaded from http://www.journalindicators.com.

- Collaboration type: Articles were categorized into three types of collaboration including individual (coded as 1), institutional (coded as 2), and international (coded as 3) collaboration. Articles written by two or more authors from the same institution are considered as individual type of collaboration. Articles published in collaboration between two or more institutions are categorized into institutional collaboration, and finally, articles published in collaboration between two or more countries are categorized into international collaboration.
- Number of references: Number of references listed in the reference list of each article was measured as a factor of citation.

- *Altmetric factors*
  - Tweet Score (*TS*): The number of times a publication is tweeted or retweeted.
  - Facebook Score (*FS*): The number of times a publication has been mentioned on Facebook wall.
  - Blog Score (*BS*): The number of times a publication has been discussed in blogs.
  - Google+ Score (*GS*): The number of times a publication has been discussed by Google+ users.
  - News Score (*NS*): The number of times a publication has been discussed in news outlets and magazines.

**Statistical Procedures**

To measure the association between citation counts and bibliometric and altmetric factors, a regression model is required. Count regression models are the best fit to the data since the dependent variable (number of citations) is a count data type. Citation data is very skewed and over-dispersed, so a standard negative binomial model is needed as it can deal with the over-dispersion. However, since the data is zero-truncated (articles with at least one citation and one altmetric count are considered), a zero-truncated negative binomial model is tested using STATA v.14.

Using advanced regression models such as a zero-truncated negative binomial model is preferred to simple correlation tests as the advanced model allows for the simultaneously examination of the association between a number of factors on citation counts, while a correlation test measures the influence of each factor on the number of citations to articles separately (Didegah et al. 2017). As noted by Thelwall et al. (2013), correlation tests alone may not be appropriate for altmetric studies as the various platforms have different levels of activity and newer publications tend to receive higher altmetric scores. The advanced model goes further than simple correlation results and calculates the percentage of increase or decrease in the citation counts for a unit change in each factor.

**Results and discussion**

This section presents the main findings of the study. First, the general presence of Almetric.com data in Scopus database is discussed. This is then followed by a discussion regarding the correlation between traditional bibliometric-based indices and web-based altmetric indices.

**What is the presence of Altmetric.com data in Scopus database overall and across the disciplines?**

This section discusses the presence of altmetric indicators in our dataset of 1,104,275 publications and describes the distribution of altmetric indicators across disciplines.

**Table 3:** Distribution of publications having different altmetric indicators

| Indicators | Publications | % within altmetrics |
| --- | --- | --- |
| Total Altmetrics | 1,104,275 | 100.00 |
| Tweets | 1,006,397 | 91.14 |
| Facebook | 245,789 | 22.26 |
| Blogs | 101,326 | 9.18 |
| News | 92,620 | 8.39 |
| Google+ | 43,709 | 3.96 |

Table 3 displays articles having received activity in online contexts; Twitter received the most activity (91.14%), followed by Facebook (22.26%), Blogs (9.18%), News (8.39%), and Google+, which demonstrated the least (3.96%). The data indicates that most of the captured activity surrounding scholarly documents occurred on Twitter, which can be regarded as the most active medium of this type of activity. Other platforms generate nominal activity with regards to scholarly documents. This is similar to the findings presented by Thelwall et al. (2013), who found that coverage in all altmetrics were relatively low, except for Twitter.

**Table 4:** Total number of publications, total citation counts (Tcc), mean citation scores (Acc), altmetric counts (FS, BS, TS, GS, NS), total number of altmetric counts (Tac), and altmetrics-to-publications ratio across discipline.

| Fields | Publications | Tcc | Acc | FS | BS | TS | GS | NS | *Tac | Tac / Publications |
|---|---|---|---|---|---|---|---|---|---|---|
| Medicine and Medical Sciences | 589429 | 7798578 | 13.23 | 430507 | 77412 | 4140183 | 45305 | 167648 | 4928761 | 8.36 |
| Biochemistry, Genetics and Molecular Biology | 294976 | 4912242 | 16.65 | 150904 | 45065 | 1571986 | 29693 | 81562 | 1929873 | 6.54 |
| Agricultural, Biological Sciences and Veterinary | 153784 | 1750640 | 11.38 | 96746 | 31145 | 965673 | 18568 | 47592 | 1185711 | 7.71 |
| Other Life and Health Sciences | 171567 | 2550107 | 14.86 | 89491 | 21989 | 810229 | 14348 | 35803 | 996083 | 5.81 |
| Social Science | 115232 | 1057757 | 9.18 | 46340 | 19419 | 677420 | 7633 | 25199 | 785207 | 6.81 |
| Health Professions and Nursing | 55327 | 565769 | 10.23 | 61232 | 5603 | 527466 | 4388 | 11669 | 614193 | 11.1 |
| Environmental Sciences | 48073 | 608935 | 12.67 | 26551 | 10068 | 246752 | 3949 | 12285 | 304187 | 6.33 |
| Engineering | 66103 | 1547611 | 23.41 | 16029 | 12315 | 193761 | 4986 | 20500 | 254329 | 3.85 |
| Chemistry | 92622 | 1799575 | 19.43 | 18953 | 10897 | 193843 | 3470 | 14797 | 247075 | 2.67 |
| Physics and Astronomy | 52402 | 1029244 | 19.64 | 13132 | 10330 | 122048 | 6289 | 21223 | 179245 | 3.42 |
| Earth and Planetary Sciences | 32033 | 452092 | 14.11 | 11530 | 11874 | 123494 | 3192 | 15303 | 171533 | 5.35 |
| Computer Science | 23288 | 327556 | 14.07 | 5312 | 2640 | 106327 | 2679 | 2072 | 121220 | 5.21 |
| Materials Sciences | 39613 | 930064 | 23.48 | 8568 | 7205 | 78538 | 3419 | 15478 | 115934 | 2.93 |
| Economics, Business and Decision Sciences | 23257 | 243305 | 10.46 | 5859 | 2725 | 89690 | 1045 | 2984 | 104325 | 4.49 |
| Mathematics | 17438 | 197378 | 11.32 | 3828 | 1967 | 82622 | 2132 | 1520 | 94248 | 5.4 |

* Table 4 is sorted by *Tac*

Table 4 represents the total number of documents present in both Altmetric.com and Scopus data and identifies mean altmetric scores and mean citations, which are categorized by discipline. The Medicine and Medical Sciences discipline was found to have more scholarly citations and a higher altmetric count than all others. Interestingly, Health Professions and Nursing shows the highest value of citation score per publication (11.1). The pattern depicted by Table 3 continues in Table 4, indicating that Twitter remains the highest social score generator across disciplines, followed by Facebook. The fields related to health, biology, agriculture, and social sciences receive greater online activity than the others, which may be indicative of greater public interest in the topics. It may also reflect differences in the ways that scholars use online environments to share and discuss science across disciplines, just as Holmberg and Thelwall (2014) found that there were clear disciplinary differences in the ways scholars used Twitter.

**How well the altmetric indices correlate with the traditional bibliometric based indices?**

This section presents the results of Spearman correlation between altmetric indicators and bibliometric indicators. The following most influential altmetric indicators were chosen for this work: Tweets Score (TS), Facebook mentions (FS), News (NS), Blogs BS), and Google+ posts (GS). In addition, the total altmetric score (Ac), citation count (Cc), and SNIP values of all publications were included for the comparative analysis. The ranks of all the selected indicators were first determined and then the correlation was tested. The confidence interval was set at 95%.

**Table 5:** Spearman correlation between altmetric and bibliometric indices

| Indicators | BS | FS | GS | NS | TS | Ac | Cc | SNIP |
|---|---|---|---|---|---|---|---|---|
| BS | 1 | .175* | .228* | .353* | .186* | .275* | .165* | .171* |
| FS |  | 1 | .190 | .185* | .248* | .391* | .085* | .119* |
| GS |  |  | 1 | .217* | .202* | .235* | .085* | .114* |
| NS |  |  |  | 1 | .197* | .297* | .125* | .187* |
| TS |  |  |  |  | 1 | .928* | .070* | .236* |
| Ac |  |  |  |  |  | 1 | .111* | .258* |
| Cc |  |  |  |  |  |  | 1 | .323* |
| SNIP |  |  |  |  |  |  |  | 1 |

*Correlation is significant at the 0.01 level (2-tailed), N = 1,104,275.

Table 5 demonstrates a positive but weak correlation between total altmetric counts and scholar citation ($\rho= 0.111$). Among the altmetric indicators, the highest correlation with citations is found for Blogs ($\rho= 0.165$). Twitter shows the most significant correlation with total altmetric counts, hence it seems to be the greatest contributing indicator, followed by Facebook mentions, News mentions, Blogs mentions, and Google+ mentions.

In addition, the correlation between total altmetric counts and scholarly citations across the disciplines was tested. The discipline Physics and Astronomy depicts the highest correlation ($\rho= 0.181$), followed by Engineering ($\rho= 0.165$), Material Science ($\rho= 0.143$), Mathematics ($\rho= 0.13$), Other Life and Health Sciences ($\rho= 0.127$), Medicine and Medical Sciences ($\rho= 0.12$), Health Professions and Nursing ($\rho= 0.116$), Biochemistry, Genetics and Molecular Biology ($\rho= 0.116$), Earth and Planetary Sciences ($\rho= 0.113$), Agricultural, Biological Sciences and Veterinary ($\rho= 0.107$), Chemistry ($\rho= 0.097$), Social Science ($\rho= 0.093$), Economics, Business and Decision Sciences ($\rho= 0.077$), Environmental Sciences ($\rho= 0.065$) and Computer Science ($\rho= 0.05$).

Table 6 represents the number of publications, their mean and standard deviation (SD), Blog score (BS), Facebook score (FS), Google+ score (GS), News score (NS), Twitter score (TS), mean of altmetric counts (Ac), average citation count (Cc), and ASNIP with respect to active altmetric indicators. The table illustrates that as the number of altmetric indicators received by publications increases, so does their average altmetric count, average citation counts, and ASNIP. It is important to realize that the number of publications naturally decreases with the increase of active altmetric indicators, since not all social platforms actively discuss all publications. It was also found that few publications receive both high altmetric count and scholarly citation.

**Table 6:** Mean and standard deviation (SD) altmetric and bibliometric indicators with respect to number of altmetrics received.

| No. of Altmetrics | Publications | Mean BS [SD] | Mean FS [SD] | Mean GS [SD] | Mean NS [SD] | Mean TS [SD] | Mean Ac [SD] | Avg Cc [SD] | *ASNIP* [SD] |
|---|---|---|---|---|---|---|---|---|---|
| 1 | 788698 | 1.14 [0.51] | 1.23 [1.51] | 1.32 [1.81] | 1.62 [1.71] | 3.07 [5.90] | 2.87 [5.65] | 11.58 [26.36] | 1.40 [1.18] |
| 2 | 217793 | 1.26 [0.66] | 1.74 [4.12] | 1.31 [2.08] | 2.06 [2.37] | 7.21 [16.51] | 8.67 [17.04] | 16.90 [42.31] | 1.84 [2.02] |
| 3 | 57019 | 1.48 [1.01] | 2.88 [21.47] | 1.50 [2.17] | 2.96 [3.49] | 16.02 [30.36] | 20.48 [43.68] | 27.89 [143.25] | 2.67 [2.90] |
| 4 | 23071 | 2.03 [1.69] | 4.58 [34.82] | 2.13 [13.19] | 4.67 [5.56] | 34.39 [103.54] | 44.33 [102.65] | 37.79 [79.84] | 3.55 [3.63] |
| 5 | 10416 | 3.03 [2.81] | 7.91 [35.66] | 3.06 [7.10] | 6.96 [8.62] | 69.23 [136.21] | 88.64 [152.05] | 57.46 [152.07] | 4.40 [4.13] |
| 6 | 4821 | 4.66 [4.40] | 14.75 [60.96] | 5.05 [10.45] | 10.06 [12.3] | 143.81 [311.86] | 179.23 [338.81] | 78.37 [186.46] | 5.16 [4.35] |
| 7 | 1768 | 7.48 [7.03] | 23.76 [77.07] | 8.13 [16.43] | 15.35 [18.72] | 205.86 [464.16] | 264.90 [517.19] | 115.12 [290.14] | 5.40 [4.38] |
| 8 | 507 | 11.64 [11.41] | 41.20 [112.58] | 15.47 [64.28] | 20.79 [24.24] | 297.16 [466.30] | 393.46 [577.87] | 156.08 [276.97] | 5.64 [4.23] |
| 9 | 145 | 17.48 [91.15] | 61.98 [113.07] | 14.43 [22.50] | 27.28 [30.30] | 465.03 [649.08] | 598.67 [785.72] | 297.46 [438.2] | 6.56 [4.26] |
| 10 | 29 | 26.72 [20.49] | 97.79 [199.5] | 30.48 [43.75] | 43.69 [43.13] | 603.24 [783.19] | 826.34 [1002.57] | 353.24 [344.14] | 7.30 [4.05] |
| 11 | 6 | 61.67 [49.58] | 117.67 [165.38] | 64.83 [53.45] | 74.17 [121.96] | 2545.83 [3932.96] | 2886.33 [4104.68] | 920.83 [142.27] | 6.71 [2.27] |
| 12 | 2 | 95.50 [120.92] | 78.00 [108.89] | 60.00 [49.50] | 41.50 [57.28] | 1367.00 [1671.60] | 1666.50 [2021.62] | 403.50 [84.15] | 5.02 [3.48] |

**Which, and to what extent, bibliometric and altmetric factors associate with number of citations to articles?**

A zero-truncated negative binomial model was run across the 15 subject fields. Similar results were found in the different fields. The results of the following seven fields with the highest number of significant factors are discussed here: Medicine & Medical Sciences, Health Professions & Nursing, Other Life & Health Sciences, Earth & Planetary Sciences, Engineering, Physics & Astronomy, and Social Sciences (see Table 7 through Table 13). Note that the results of the model for other fields are presented in the Appendix A, Table A-1 through Table A-8.

In Medicine & Medical Sciences, the results of the zero-truncated NB model indicate that all factors in the field are significantly positively associated with citation counts. Similar results were obtained in Health Professions & Nursing, Other Life & Health Sciences, and Physics & Astronomy. However, in Earth & Planetary Sciences, Engineering, and Social Science news counts depict a weak negative association with citations counts.

Among the bibliometric factors, document type is most strongly associated with increased citations in Medicine & Medical Sciences; reviews receive 27.8% higher citations than the letters, articles, and non-citation documents. A similar behavior is observed in other fields.

Journal prestige, as measured by SNIP, is another significant factor. A unit change in the SNIP increases the number of citations by 13.9%, 30.3%, 68%, 45.1%, 53.8% in Medicine & Medical Sciences, Other Life & Health Sciences, Earth & Planetary Sciences, Physics & Astronomy, and Social Sciences, respectively. Interestingly, Engineering shows the highest significance (i.e. 143.9%), followed by Health Professions & Nursing with 121.1% in this regard. Journal impact, typically measured by the Journal Impact Factor, was also found to be the most important determinant of citations in previous literature (Didegah and Thelwall 2013; Boyack and Klavans 2005).

Type of collaboration is also a significant determinant of citations. The positive association illustrates that articles involved in an international collaboration receive more citations than articles of individual and institutional types. Among all the fields, the maximum contribution (i.e. 18.5%) of international collaborations towards citation counts has been observed by Earth & Planetary Sciences, Biochemistry, and Genetics & Molecular Biology. International collaboration was also an important factor for citations as reported in previous works (Didegah and Thelwall 2013). International collaboration is the widest type of collaboration through which authors from different institutions and different countries get involved in research. This wide type of collaboration increases the chance of article visibility globally, which may result in more citations later.

Finally, the number of references significantly associates with increased citation counts. However, the extent to which this factor associates with the number of citations is weak; the number of citations increases by only 0.9% for one more reference in the reference list in the field of Health Professions & Nursing, which is the maximum increase across all the fields.

Among altmetric indicators, Blog count is very strongly contributing to increased citations. According to the results, one more blog post discussing an article increases the chance of more citations by 36.8% in Health Professions & Nursing – the maximum by any field. Looking at other altmetric indicators, one additional tweet, news post, FB post, or Google+ post about a paper in Medicine & Medical Sciences increased the likelihood of more citations by 1.7%, 6%, 6.3%, and 8.9%, respectively. While in Health Professions & Nursing, one more Google+ post about the paper increased the likelihood of more citations by 9.6%. In Other Life & Health Sciences, one more news post about the paper increased the likelihood of more citations by 2.5%. In Physics & Astronomy, tweet counts contributed to increased citations by 7.7%. Interestingly, news counts negatively contribute to citations in the following fields: Earth & Planetary Sciences, Engineering, and Social Science.

Table 7: The results of zero-truncated NB model in Medicine & Medical Sciences

| scopus_citation | Coef. | Exp.(Coef.) | Std. Err. | z | P>z | [95% Conf. | Interval] |
|---|---|---|---|---|---|---|---|
| SNIP | 0.130 | 1.139 | 0.004 | 32.17 | 0.000 | 0.122 | 0.138 |
| Doc. type | 0.246 | 1.278 | 0.037 | 6.69 | 0.000 | 0.174 | 0.318 |
| Collab. type | 0.099 | 1.104 | 0.025 | 4.01 | 0.000 | 0.050 | 0.147 |
| No. refs | 0.005 | 1.005 | 0.000 | 11.34 | 0.000 | 0.004 | 0.005 |
| Tweet count | 0.017 | 1.017 | 0.000 | 98.24 | 0.000 | 0.017 | 0.017 |
| News count | 0.058 | 1.060 | 0.002 | 30.97 | 0.000 | 0.054 | 0.062 |
| Google+ count | 0.086 | 1.089 | 0.007 | 13.03 | 0.000 | 0.073 | 0.099 |
| FB count | 0.061 | 1.063 | 0.001 | 45.66 | 0.000 | 0.059 | 0.064 |
| Blog count | 0.221 | 1.247 | 0.005 | 41.07 | 0.000 | 0.210 | 0.232 |

Table 8: The results of zero-truncated NB model in Health Professions & Nursing

| scopus_citation | Coef. | Exp.(Coef.) | Std. Err. | z | P>z | [95% Conf. | Interval] |
|---|---|---|---|---|---|---|---|
| SNIP | 0.793 | 2.211 | 0.011 | 71.810 | 0.000 | 0.772 | 0.815 |
| Doc. type | 0.183 | 1.200 | 0.010 | 18.480 | 0.000 | 0.163 | 0.202 |
| Collab. type | 0.070 | 1.073 | 0.008 | 8.740 | 0.000 | 0.054 | 0.086 |
| No. refs | 0.009 | 1.009 | 0.000 | 33.910 | 0.000 | 0.008 | 0.009 |
| Tweet count | 0.010 | 1.010 | 0.000 | 26.740 | 0.000 | 0.009 | 0.011 |
| News count | 0.033 | 1.034 | 0.007 | 4.660 | 0.000 | 0.019 | 0.047 |
| Google+ count | 0.092 | 1.096 | 0.023 | 3.980 | 0.000 | 0.047 | 0.137 |
| FB count | 0.043 | 1.044 | 0.002 | 17.590 | 0.000 | 0.038 | 0.048 |
| Blog count | 0.313 | 1.368 | 0.025 | 12.660 | 0.000 | 0.265 | 0.362 |

Table 9: The results of zero-truncated NB model in Other Life & Health Sciences

| scopus_citation | Coef. | Exp.(Coef.) | Std. Err. | z | P>z | [95% Conf. | Interval] |
|---|---|---|---|---|---|---|---|
| SNIP | 0.265 | 1.303 | 0.024 | 11.110 | 0.000 | 0.218 | 0.311 |
| Doc. type | 0.223 | 1.249 | 0.060 | 3.690 | 0.000 | 0.104 | 0.341 |
| Collab. type | 0.125 | 1.133 | 0.039 | 3.180 | 0.001 | 0.048 | 0.202 |
| No. refs | 0.002 | 1.002 | 0.001 | 2.460 | 0.014 | 0.000 | 0.003 |
| Tweet count | 0.021 | 1.021 | 0.001 | 41.950 | 0.000 | 0.020 | 0.022 |
| News count | 0.025 | 1.026 | 0.004 | 6.580 | 0.000 | 0.018 | 0.033 |
| Google+ count | 0.017 | 1.017 | 0.005 | 3.810 | 0.000 | 0.008 | 0.026 |
| FB count | 0.007 | 1.007 | 0.002 | 4.330 | 0.000 | 0.004 | 0.009 |
| Blog count | 0.099 | 1.104 | 0.008 | 12.420 | 0.000 | 0.083 | 0.114 |

Table 10: The results of zero-truncated NB model in Earth & Planetary Sciences

| scopus_citation | Coef. | Exp.(Coef.) | Std. Err. | z | P>z | [95% Conf. | Interval] |
|---|---|---|---|---|---|---|---|
| SNIP | 0.519 | 1.680 | 0.016 | 32.810 | 0.000 | 0.488 | 0.550 |
| Doc. type | 0.067 | 1.070 | 0.022 | 3.040 | 0.002 | 0.024 | 0.111 |
| Collab. type | 0.170 | 1.185 | 0.009 | 18.730 | 0.000 | 0.152 | 0.188 |
| No. refs | 0.006 | 1.006 | 0.000 | 28.760 | 0.000 | 0.006 | 0.006 |
| Tweet count | 0.016 | 1.016 | 0.001 | 15.650 | 0.000 | 0.014 | 0.018 |
| News count | -0.016 | 0.984 | 0.006 | -2.720 | 0.007 | -0.028 | -0.004 |
| Google+ count | -0.001 | 0.999 | 0.010 | -0.080 | 0.937 | -0.021 | 0.020 |
| FB count | 0.041 | 1.042 | 0.008 | 5.480 | 0.000 | 0.027 | 0.056 |
| Blog count | 0.109 | 1.115 | 0.008 | 14.420 | 0.000 | 0.094 | 0.124 |

Table 11: The results of zero-truncated NB model in Engineering

| scopus_citation | Coef. | Exp.(Coef.) | Std. Err. | z | P>z | [95% Conf. | Interval] |
|---|---|---|---|---|---|---|---|
| SNIP | 0.892 | 2.439 | 0.013 | 68.060 | 0.000 | 0.866 | 0.917 |
| Doc. type | 0.245 | 1.277 | 0.013 | 18.910 | 0.000 | 0.219 | 0.270 |
| Collab. type | 0.066 | 1.069 | 0.007 | 9.110 | 0.000 | 0.052 | 0.081 |
| No. refs | 0.005 | 1.005 | 0.000 | 24.980 | 0.000 | 0.005 | 0.006 |
| Tweet count | 0.008 | 1.008 | 0.001 | 13.090 | 0.000 | 0.007 | 0.009 |
| News count | -0.022 | 0.978 | 0.011 | -2.020 | 0.043 | -0.043 | -0.001 |
| Google+ count | 0.033 | 1.033 | 0.009 | 3.450 | 0.001 | 0.014 | 0.051 |
| FB count | -0.001 | 0.999 | 0.001 | -1.240 | 0.216 | -0.002 | 0.000 |
| Blog count | 0.124 | 1.132 | 0.011 | 11.290 | 0.000 | 0.102 | 0.145 |

**Table 12**: The results of zero-truncated NB model in Physics & Astronomy

| scopus_citation | Coef. | Exp.(Coef.) | Std. Err. | z | P>z | [95% Conf. | Interval] |
|---|---|---|---|---|---|---|---|
| SNIP | 0.373 | 1.451 | 0.005 | 67.870 | 0.000 | 0.362 | 0.383 |
| Doc. type | 0.248 | 1.281 | 0.020 | 12.400 | 0.000 | 0.209 | 0.287 |
| Collab. type | 0.145 | 1.156 | 0.008 | 19.300 | 0.000 | 0.130 | 0.159 |
| No. refs | 0.006 | 1.006 | 0.000 | 28.040 | 0.000 | 0.005 | 0.006 |
| Tweet count | 0.077 | 1.080 | 0.003 | 24.890 | 0.000 | 0.071 | 0.083 |
| News count | 0.015 | 1.015 | 0.004 | 3.580 | 0.000 | 0.007 | 0.023 |
| Google+ count | 0.003 | 1.003 | 0.007 | 0.500 | 0.620 | -0.010 | 0.017 |
| FB count | 0.047 | 1.048 | 0.014 | 3.270 | 0.001 | 0.019 | 0.075 |
| Blog count | 0.093 | 1.098 | 0.013 | 6.980 | 0.000 | 0.067 | 0.120 |

**Table 13**: The results of zero-truncated NB model in Social Sciences

| scopus_citation | Coef. | Exp.(Coef.) | Std. Err. | z | P>z | [95% Conf. | Interval] |
|---|---|---|---|---|---|---|---|
| SNIP | 0.431 | 1.538 | 0.007 | 65.830 | 0.000 | 0.418 | 0.443 |
| Doc. type | 0.237 | 1.267 | 0.010 | 23.350 | 0.000 | 0.217 | 0.256 |
| Collab. type | 0.151 | 1.163 | 0.006 | 26.730 | 0.000 | 0.140 | 0.162 |
| No. refs | 0.007 | 1.007 | 0.000 | 42.950 | 0.000 | 0.006 | 0.007 |
| Tweet count | 0.013 | 1.013 | 0.000 | 26.300 | 0.000 | 0.012 | 0.014 |
| News count | -0.002 | 0.998 | 0.004 | -0.580 | 0.565 | -0.009 | 0.005 |
| Google+ count | 0.085 | 1.089 | 0.017 | 5.030 | 0.000 | 0.052 | 0.118 |
| FB count | 0.065 | 1.068 | 0.006 | 11.810 | 0.000 | 0.054 | 0.076 |
| Blog count | 0.157 | 1.170 | 0.010 | 16.390 | 0.000 | 0.138 | 0.176 |

When examined in a single table (as shown in Table 14), the results indicate the differences between the factors and all fields. As indicated in the table and the discussion above, the field of Health Professions and Nursing had strong positive scores across all factors including the highest increase in citations with number of references (0.9%), news postings (3.4%), and blog postings (36.8%). The field of Chemistry demonstrates the most negative scores with a decrease in citation numbers associated with collaboration (-1.7%), news postings (-2.9%), and Google+ postings (-1.1%).

**Table 14:** The results of zero-truncated NB model in all fields shown as percentages of increase (or decrease) in citations related to factors. TC = Twitter Count, NC = News count, GC = Google+ count, FBC = Facebook count; BC = Blog count: **bold** indicates highest positive score in factor; *italics* indicates lowest score in factor.

| Field | SNIP | Doc. Type | Collab. Type | No. Refs | TC | NC | GC | FBC | BC |
|---|---|---|---|---|---|---|---|---|---|
| Agricultural, Biological Sciences and Veterinary | 48 | **32.8** | 7.9 | 0.3 | 0.7 | 0.6 | -0.2 | 0.1 | 7.3 |
| Biochemistry, Genetics and Molecular Biology | 46.2 | 30.6 | 6.3 | *0.2* | 1.8 | 1.5 | -0.1 | -0.2 | 8 |
| Chemistry | 70.2 | 7.9 | *-1.7* | 0.5 | 3.8 | *-2.9* | *-1.1* | 0 | 12.6 |
| Computer Science | 76.1 | 20.8 | 11.9 | 0.7 | 2.2 | -1.4 | 1.7 | -0.4 | 18.7 |
| Earth and Planetary Sciences | 68 | *7* | **18.5** | 0.6 | 1.6 | 1.5 | -0.1 | -0.2 | 8 |
| Economics, Business and Decision Sciences | 49.5 | 11.5 | 8.2 | 0.7 | 1.5 | -2.6 | 2.1 | 3.8 | 24.7 |
| Engineering | **143.9** | 27.7 | 6.9 | 0.5 | 0.8 | -2.2 | 3.3 | -0.1 | 13.2 |
| Environmental Sciences | 45 | 31.1 | 7.4 | 0.3 | 0.7 | -1.5 | -0.2 | 0.1 | *6.9* |
| Health Professions and Nursing | 121.1 | 20 | 7.3 | **0.9** | 1 | **3.4** | 9.6 | 4.4 | **36.8** |
| Materials Sciences | 48 | 13.3 | 1.5 | 0.7 | 7 | -1.9 | -0.4 | **9.6** | 15.9 |
| Mathematics | 76.5 | 26.5 | 13.4 | 0.8 | 3.7 | -0.4 | **14.9** | *-1.3* | 17.2 |
| Medicine and Medical Sciences | *13.9* | 27.8 | 10.4 | 0.5 | 1.7 | 6 | 8.9 | 6.3 | 24.7 |
| Other Life and Health Sciences | 30.3 | 24.9 | 13.3 | *0.2* | 2.1 | 2.6 | 1.7 | 0.7 | 10.4 |
| Physics and Astronomy | 45.1 | 28.1 | 15.6 | 0.6 | **8** | 1.5 | 0.3 | 4.8 | 9.8 |
| Social Sciences | 53.8 | 26.7 | 16.3 | 0.7 | 1.3 | -0.2 | 8.9 | 6.8 | 17 |

**Can altmetric scores discriminate publications with a higher citation impact?**

This section presents the discussion on altmetric scores, through discriminating the HC (high citation) 1% papers and utilizing a receiver operating characteristic (ROC) curve. The ROC is a graphical plot that illustrates the performance of a binary classifier system as its discrimination threshold is varied. The curve is created by plotting the true positive rate (TPR) against the false positive rate (FPR) at various threshold settings.

The behavior of SNIP and scholarly citations is examined to discriminate the top 1% highly-cited publications.

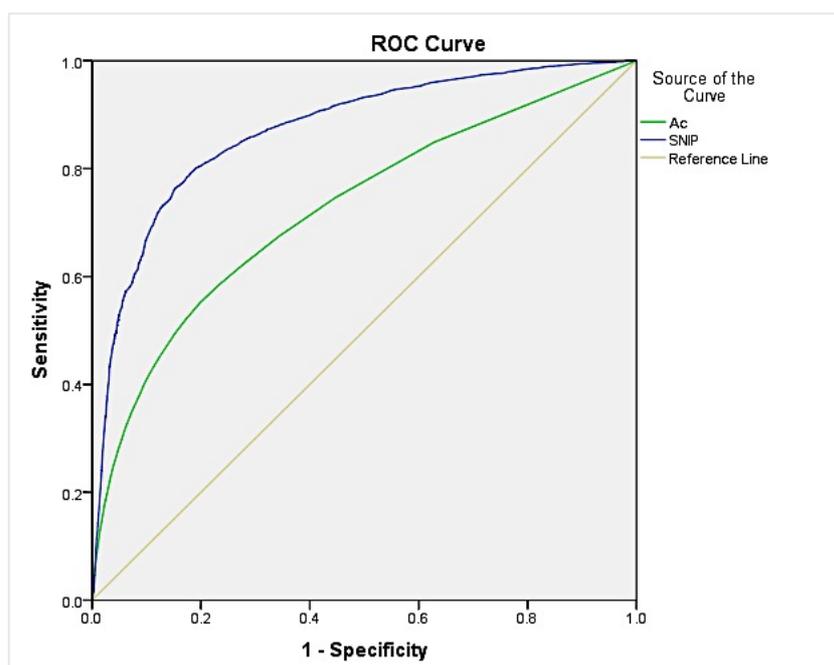

**Figure 2:** ROC curve for SNIP and Total Altmetric counts (Ac) to identify HC 1% publications

Figure 2 shows the area under the curve for SNIP and total altmetric counts to discriminate HC 1% publications. Since SNIP is a citation based indicator, it is natural to achieve a very promising area under the curve (AUC = 0.871). Interestingly, the results demonstrate an encouraging area under the curve (AUC = 0.725) for total altmetric counts as well. This analysis indicates that total altmetric counts could be used as a notable indicator to discriminate highly cited publications.

Further, the ROC curve is presented across disciplines (see Table 15). It was determined that the Medicine and Medical Science (AUC= 0.736), Physics & Astronomy (AUC= 0.726), and Other Life & Health Sciences (AUC= 0.719) demonstrate very encouraging results, with all disciplines exhibiting at least 0.7 areas under the curve. Interestingly, Computer Science shows an even higher area under the curve for total altmetric counts as compared to SNIP. These results indicate that total altmetric counts are an even better indicator of discriminating highly cited papers as compared to SNIP for the discipline of Computer Sciences. The ROC curves for each discipline are given in Appendix B Fig. B-1 through Fig. B-3.

**Table 15:** ROC curve for SNIP and Total Altmetric counts (Ac) to identify HC 1% publications across discipline

|  | AUC | |
|---|---|---|
| **Fields** | Ac | SNIP |
| Medicine and Medical Sciences | 0.736 | 0.854 |
| Physics & Astronomy | 0.726 | 0.868 |
| Other Life & Health Sciences | 0.719 | 0.847 |
| Materials Science | 0.698 | 0.885 |
| Computer Science | 0.67 | 0.626 |
| Engineering | 0.666 | 0.864 |
| Biochemistry, Genetics and Molecular Biology | 0.656 | 0.796 |
| Social Sciences | 0.647 | 0.738 |
| Mathematics | 0.645 | 0.739 |
| Economics, Business & Decision Sciences | 0.632 | 0.788 |
| Health Professions & Nursing | 0.632 | 0.815 |
| Earth and Planetary Sciences | 0.623 | 0.655 |
| Chemistry | 0.621 | 0.88 |
| Environmental Science | 0.601 | 0.733 |
| Agricultural, Biological Sciences & Veterinary | 0.592 | 0.67 |

**Which altmetric indicator is most significant to discriminate highly cited publications?**

This section presents the discussion on the ability of altmetric indicators to discriminate highly cited publications. The ROC curves of the top five altmetric indicators are presented for the HC 1% papers, with the goal to measure which indicator best distinguishes highly cited papers (see Fig. 3). It was found that Blog scores outperform all other indicators, followed by the Twitter score suggesting that Blog posts and Twitter posts are the most active mediums used to communicate and share publications, especially in the case of highly cited papers. This is demonstrated by the observation that all highly cited papers were discussed in Blog posts in the dataset. The importance of blogs for increasing citation counts was reported earlier by Shema et al. (2014).

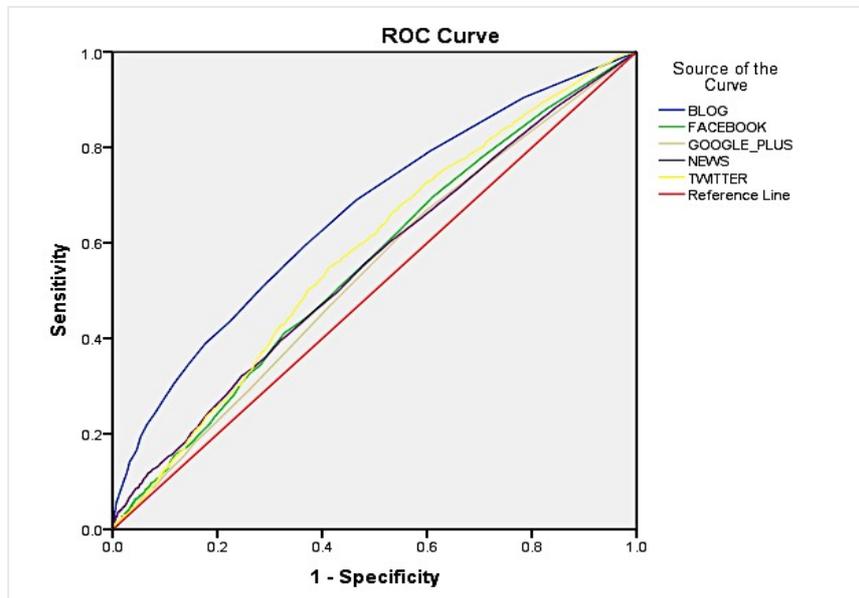

**Figure 3:** ROC of top five altmetric indicators against the HC 1% papers

**Concluding remarks**

In this paper, the authors have examined the online activity of scholarly articles across all broader scientific disciplines indexed in the Scopus database and captured by Altmetric.com. It was found that there is a rapid increase in the coverage of Altmetric.com data present in the Scopus database, with 20.46% increase in 2015 from 10.19% in 2011. This is similar to previous research (Haustein et al. 2014a), which also showed an increase in captured Altmetric events across time. The results have also shown that Twitter has the most significant impact among social indicators (with 91.14% presence), followed by Facebook (with 22.26%). Again, this is similar to previous research (Thelwall et al. 2013) that has shown Twitter to be the most active social media platform for sharing scientific works.

Despite having a weak positive correlation between bibliometric and altmetric indices, altmetric indices could be a useful indicator to differentiate publications with citation impact. With regards to different ASJC disciplines, results found that research in health, biology, agriculture, and social sciences receive higher amounts of activity, which may be indicative of greater public interest in these topics. The Medicine and Medical Sciences received more citations and higher total altmetric counts scores than all others; the highest number of citations was found in the field of Health Professions and Nursing. A higher altmetric activity around publications may be reflective of a greater public interest in these publications compared to the other publications in the field.

Of particular interest was that all highly-cited papers were discussed in Blog posts in the dataset analyzed. It was also found that Blog count is very strongly contributing to an increase in citation counts. These two results demonstrate the importance of monitoring blog posts written on articles, which could be beneficial for scholars to utilize them as filters to determine the most impactful articles. Overall, the findings suggest that altmetrics could be used to better distinguish highly cited publications. However, additional research is required to reveal the semantic understanding of altmetric indices for research evaluation (Liu and Fang 2017).

Finally, there were some limitations with the data used in this study. Altmetric.com primarily captures articles having a DOI, therefore not all the publications indexed in Scopus have been associated with activity in social media contexts. Another limitation was the presence of corrupt DOIs in Altmetric.com dataset, which could not be matched with Scopus.

**Acknowledgements**

We are thankful to Almetric.com for providing the dataset.

**References**

Adie, E., & Roe, W. (2013). Altmetric: Enriching scholarly content with article-level discussion and metrics. *Learned Publishing*, *26*(1), 11-17.

Bar-Ilan, J., Haustein, S., Peters, I., Priem, J., Shema, H., & Terliesner, J. (2012). Beyond citations: Scholars' visibility on the social Web. *arXiv preprint arXiv:1205.5611*.

Boyack, K. W. & Klavans, R. (2005). Predicting the importance of current papers. In P. Ingwersen and B. Larsen (Ed.) *Proceedings of ISSI 2005* (pp. 335–342). Stockholm, Sweden.

Brody, T., Harnad, S., & Carr, L. (2006). Earlier web usage statistics as predictors of later citation impact. *Journal of the Association for Information Science and Technology*, 57(8), 1060-1072.

Costas, R., Zahedi, Z., & Wouters, P. (2015). Do "altmetrics" correlate with citations? Extensive comparison of altmetric indicators with citations from a multidisciplinary perspective. *Journal of the Association for Information Science and Technology*, *66*(10), 2003-2019.

Didegah, F., & Thelwall, M. (2013). Which factors help authors produce the highest impact research? Collaboration, journal and document properties. Journal of Informetrics, 7(4), 861-873.

Didegah, F., Bowman, T.D., & Holmberg, K. (2017). On the differences between citations and altmetrics: An investigation of factors driving altmetrics vs. citations. *Journal of the Association for information Science and Technology,* (in press).

Haddawy, P., Hassan, S. U., Abbey, C. W., & Lee, I. B. (2017). Uncovering fine-grained research excellence: The global research benchmarking system. *Journal of Informetrics*, *11*(2), 389-406.

Hassan, S. U., & Gillani, U. A. (2016). Altmetrics of" altmetrics" using Google Scholar, Twitter, Mendeley, Facebook, Google-plus, CiteULike, Blogs and Wiki. *arXiv preprint arXiv:1603.07992*.

Haustein, S., & Siebenlist, T. (2011). Applying social bookmarking data to evaluate journal usage. *Journal of informetrics*, *5*(3), 446-457.

Haustein, S., Peters, I., Sugimoto, C. R., Thelwall, M., & Larivière, V. (2014a). Tweeting biomedicine: An analysis of tweets and citations in the biomedical literature. *Journal of the Association for Information Science and Technology*, *65*(4), 656-669.

Haustein, S., Peters, I., Bar-Ilan, J., Priem, J., Shema, H., & Terliesner, J. (2014b). Coverage and adoption of altmetrics sources in the bibliometric community. *Scientometrics*, *101*(2), 1145-1163.

Holmberg, K., & Thelwall, M. (2014). Disciplinary differences in Twitter scholarly communication. *Scientometrics, 101*(2), 1027–1042. doi: 10.1007/s11192-014-1229-3

Liu, X. Z., & Fang, H. (2017). What we can learn from tweets linking to research papers. *Scientometrics*, *111*(1), 349-369.

Nielsen, F. (2007). Scientific citations in Wikipedia. *arXiv preprint arXiv:0705.2106*.

Priem, J., & Hemminger, B. H. (2010). Scientometrics 2.0: New metrics of scholarly impact on the social Web. *First Monday*, *15*(7).

Priem, J., Piwowar, H. A., & Hemminger, B. M. (2012). Altmetrics in the wild: Using social media to explore scholarly impact. *arXiv preprint arXiv:1203.4745*.

Priem, J., Taraborelli, D., Groth, P., & Neylon, C. (2010). Altmetrics: A manifesto. Available at: http://altmetrics.org/manifesto/

Shema H, Bar-Ilan J, Thelwall M. (2014). Do blog citations correlate with a higher number of future citations? Research blogs as a potential source for alternative metrics. *Journal of the Association for Information Science and Technology. 65*(5):1018-27

Sud, P., & Thelwall, M. (2014). Evaluating altmetrics. *Scientometrics*, *98*(2), 1131-1143.

Sugimoto, C. R., Russell, T. G., Meho, L. I., & Marchionini, G. (2008). MPACT and citation impact: Two sides of the same scholarly coin? *Library & Information Science Research*, *30*(4), 273-281.

Sugimoto, C. R., Work, S., Larivière, V., & Haustein, S. (2017). Scholarly use of social media and altmetrics: A review of the literature. *Journal of the Association for Information Science and Technology*. doi: 10.1002/asi.23833

Thelwall, M., Haustein, S., Larivière, V., & Sugimoto, C. R. (2013). Do altmetrics work? Twitter and ten other social web services. *PloS one*, *8*(5), e64841.

Wouters, P., & Costas, R. (2012). *Users, narcissism and control: tracking the impact of scholarly publications in the 21st century* (pp. 847-857). Utrecht: SURFfoundation.

Yu, H. (2017). Context of altmetrics data matters: An investigation of count type and user category. *Scientometrics, 111*(1), 267-283.

Zahedi, Z., Costas, R., & Wouters, P. (2013, October). What is the impact of the publications read by the different mendeley users? Could they help to identify alternative types of impact? Paper presented at the PLoS ALM Workshop, San Francisco, CA.

Zahedi, Z., Costas, R., & Wouters, P. (2014). How well developed are altmetrics? A cross-disciplinary analysis of the presence of 'alternative metrics' in scientific publications. *Scientometrics*, *101*(2), 1491-1513.

# Appendix A

**Table A-1:** The results of zero-truncated NB model in Agricultural, Biological Sciences and Veterinary

| scopus_citation | Coef. | Exp.(Coef.) | Std. Err. | z | P>z | [95% Conf. | Interval] |
|---|---|---|---|---|---|---|---|
| SNIP | 0.392 | 1.480 | 0.070 | 5.580 | 0.000 | 0.255 | 0.530 |
| Doc. type | 0.284 | 1.328 | 0.074 | 3.830 | 0.000 | 0.139 | 0.429 |
| Collab. type | 0.076 | 1.079 | 0.036 | 2.120 | 0.034 | 0.006 | 0.147 |
| No. refs | 0.003 | 1.003 | 0.001 | 3.660 | 0.000 | 0.001 | 0.004 |
| Tweet count | 0.007 | 1.007 | 0.000 | 26.220 | 0.000 | 0.006 | 0.008 |
| News count | 0.006 | 1.006 | 0.002 | 3.300 | 0.001 | 0.002 | 0.009 |
| Google+ count | -0.002 | 0.998 | 0.001 | -1.360 | 0.173 | -0.004 | 0.001 |
| FB count | 0.001 | 1.001 | 0.001 | 1.560 | 0.119 | 0.000 | 0.003 |
| Blog count | 0.070 | 1.073 | 0.008 | 9.250 | 0.000 | 0.055 | 0.085 |

**Table A-2:** The results of zero-truncated NB model in Biochemistry, Genetics and Molecular Biology

| scopus_citation | Coef. | Exp.(Coef.) | Std. Err. | z | P>z | [95% Conf. | Interval] |
|---|---|---|---|---|---|---|---|
| SNIP | 0.380 | 1.462 | 0.015 | 25.480 | 0.000 | 0.351 | 0.409 |
| Doc. type | 0.267 | 1.306 | 0.058 | 4.570 | 0.000 | 0.153 | 0.382 |
| Collab. type | 0.062 | 1.063 | 0.031 | 2.010 | 0.044 | 0.002 | 0.121 |
| No. refs | 0.002 | 1.002 | 0.001 | 2.750 | 0.006 | 0.001 | 0.003 |
| Tweet count | 0.018 | 1.018 | 0.000 | 55.720 | 0.000 | 0.017 | 0.018 |
| News count | 0.015 | 1.015 | 0.004 | 3.970 | 0.000 | 0.008 | 0.022 |
| Google+ count | -0.001 | 0.999 | 0.010 | -0.080 | 0.937 | -0.021 | 0.020 |
| FB count | -0.002 | 0.998 | 0.002 | -0.930 | 0.353 | -0.007 | 0.003 |
| Blog count | 0.077 | 1.080 | 0.011 | 6.760 | 0.000 | 0.055 | 0.099 |

**Table A-3:** The results of zero-truncated NB model in Chemistry

| scopus_citation | Coef. | Exp.(Coef.) | Std. Err. | z | P>z | [95% Conf. | Interval] |
|---|---|---|---|---|---|---|---|
| SNIP | 0.532 | 1.702 | 0.005 | 98.280 | 0.000 | 0.521 | 0.542 |
| Doc. type | 0.076 | 1.079 | 0.010 | 7.790 | 0.000 | 0.057 | 0.095 |
| Collab. type | -0.017 | 0.983 | 0.005 | -3.560 | 0.000 | -0.027 | -0.008 |
| No. refs | 0.005 | 1.005 | 0.000 | 38.460 | 0.000 | 0.005 | 0.005 |
| Tweet count | 0.037 | 1.038 | 0.002 | 21.590 | 0.000 | 0.034 | 0.040 |
| News count | -0.030 | 0.971 | 0.014 | -2.130 | 0.033 | -0.057 | -0.002 |
| Google+ count | -0.011 | 0.989 | 0.007 | -1.500 | 0.134 | -0.026 | 0.003 |
| FB count | 0.000 | 1.000 | 0.002 | 0.100 | 0.923 | -0.004 | 0.004 |
| Blog count | 0.119 | 1.126 | 0.015 | 7.940 | 0.000 | 0.089 | 0.148 |

**Table A-4:** The results of zero-truncated NB model in Computer Science

| scopus_citation | Coef. | Exp.(Coef.) | Std. Err. | z | P>z | [95% Conf. | Interval] |
|---|---|---|---|---|---|---|---|
| SNIP | 0.566 | 1.761 | 0.183 | 3.100 | 0.002 | 0.208 | 0.924 |
| Doc. type | 0.189 | 1.208 | 0.031 | 6.040 | 0.000 | 0.127 | 0.250 |
| Collab. type | 0.112 | 1.119 | 0.014 | 8.240 | 0.000 | 0.086 | 0.139 |
| No. refs | 0.007 | 1.007 | 0.000 | 19.570 | 0.000 | 0.007 | 0.008 |
| Tweet count | 0.022 | 1.022 | 0.001 | 14.620 | 0.000 | 0.019 | 0.025 |
| News count | -0.014 | 0.986 | 0.017 | -0.840 | 0.400 | -0.047 | 0.019 |
| Google+ count | 0.016 | 1.017 | 0.016 | 1.030 | 0.304 | -0.015 | 0.048 |
| FB count | -0.004 | 0.996 | 0.014 | -0.320 | 0.750 | -0.032 | 0.023 |
| Blog count | 0.172 | 1.187 | 0.031 | 5.500 | 0.000 | 0.110 | 0.233 |

**Table A-5:** The results of zero-truncated NB model in Economics, Business & Decision Sciences

| scopus_citation | Coef. | Exp.(Coef.) | Std. Err. | z | P>z | [95% Conf. | Interval] |
|---|---|---|---|---|---|---|---|
| SNIP | 0.402 | 1.495 | 0.011 | 36.940 | 0.000 | 0.381 | 0.423 |
| Doc. type | 0.109 | 1.115 | 0.028 | 3.860 | 0.000 | 0.053 | 0.164 |
| Collab. type | 0.078 | 1.082 | 0.011 | 6.950 | 0.000 | 0.056 | 0.101 |
| No. refs | 0.007 | 1.007 | 0.000 | 23.190 | 0.000 | 0.007 | 0.008 |
| Tweet count | 0.014 | 1.015 | 0.001 | 9.690 | 0.000 | 0.012 | 0.017 |
| News count | -0.026 | 0.974 | 0.021 | -1.230 | 0.218 | -0.068 | 0.016 |
| Google+ count | 0.021 | 1.021 | 0.028 | 0.760 | 0.446 | -0.033 | 0.076 |
| FB count | 0.038 | 1.038 | 0.013 | 2.920 | 0.003 | 0.012 | 0.063 |
| Blog count | 0.221 | 1.247 | 0.035 | 6.230 | 0.000 | 0.151 | 0.290 |

**Table A-6:** The results of zero-truncated NB model in Environmental Science

| scopus_citation | Coef. | Exp.(Coef.) | Std. Err. | z | P>z | [95% Conf. | Interval] |
|---|---|---|---|---|---|---|---|
| SNIP | 0.372 | 1.450 | 0.061 | 6.070 | 0.000 | 0.252 | 0.491 |
| Doc. type | 0.271 | 1.311 | 0.065 | 4.170 | 0.000 | 0.144 | 0.398 |
| Collab. type | 0.072 | 1.074 | 0.032 | 2.260 | 0.024 | 0.010 | 0.134 |
| No. refs | 0.003 | 1.003 | 0.001 | 4.060 | 0.000 | 0.001 | 0.004 |
| Tweet count | 0.007 | 1.007 | 0.000 | 26.220 | 0.000 | 0.006 | 0.008 |
| News count | -0.015 | 0.985 | 0.002 | -7.310 | 0.000 | -0.019 | -0.011 |
| Google+ count | -0.002 | 0.998 | 0.001 | -1.370 | 0.170 | -0.004 | 0.001 |
| FB count | 0.001 | 1.001 | 0.001 | 1.640 | 0.101 | 0.000 | 0.002 |
| Blog count | 0.067 | 1.069 | 0.007 | 10.080 | 0.000 | 0.054 | 0.080 |

**Table A-7:** The results of zero-truncated NB model in Materials Science

| scopus_citation | Coef. | Exp.(Coef.) | Std. Err. | z | P>z | [95% Conf. | Interval] |
|---|---|---|---|---|---|---|---|
| SNIP | 0.392 | 1.480 | 0.005 | 71.990 | 0.000 | 0.381 | 0.403 |
| Doc. type | 0.125 | 1.133 | 0.020 | 6.290 | 0.000 | 0.086 | 0.164 |
| Collab. type | 0.015 | 1.015 | 0.008 | 1.890 | 0.059 | -0.001 | 0.030 |
| No. refs | 0.007 | 1.007 | 0.000 | 29.220 | 0.000 | 0.007 | 0.008 |
| Tweet count | 0.067 | 1.070 | 0.003 | 20.080 | 0.000 | 0.061 | 0.074 |
| News count | -0.019 | 0.981 | 0.011 | -1.740 | 0.083 | -0.040 | 0.002 |
| Google+ count | -0.004 | 0.996 | 0.007 | -0.560 | 0.577 | -0.018 | 0.010 |
| FB count | 0.091 | 1.096 | 0.018 | 5.170 | 0.000 | 0.057 | 0.126 |
| Blog count | 0.147 | 1.159 | 0.018 | 8.270 | 0.000 | 0.112 | 0.182 |

**Table A-8:** The results of zero-truncated NB model in Mathematics

| scopus_citation | Coef. | Exp.(Coef.) | Std. Err. | z | P>z | [95% Conf. | Interval] |
|---|---|---|---|---|---|---|---|
| SNIP | 0.568 | 1.765 | 0.027 | 20.660 | 0.000 | 0.514 | 0.622 |
| Doc. type | 0.235 | 1.265 | 0.047 | 4.950 | 0.000 | 0.142 | 0.328 |
| Collab. type | 0.126 | 1.134 | 0.017 | 7.600 | 0.000 | 0.093 | 0.158 |
| No. refs | 0.008 | 1.008 | 0.001 | 13.330 | 0.000 | 0.007 | 0.009 |
| Tweet count | 0.036 | 1.037 | 0.002 | 16.030 | 0.000 | 0.032 | 0.040 |
| News count | -0.004 | 0.996 | 0.019 | -0.240 | 0.811 | -0.041 | 0.032 |
| Google+ count | 0.139 | 1.149 | 0.051 | 2.730 | 0.006 | 0.039 | 0.239 |
| FB count | -0.013 | 0.987 | 0.015 | -0.820 | 0.410 | -0.043 | 0.018 |
| Blog count | 0.159 | 1.172 | 0.050 | 3.170 | 0.002 | 0.061 | 0.257 |

# Appendix B

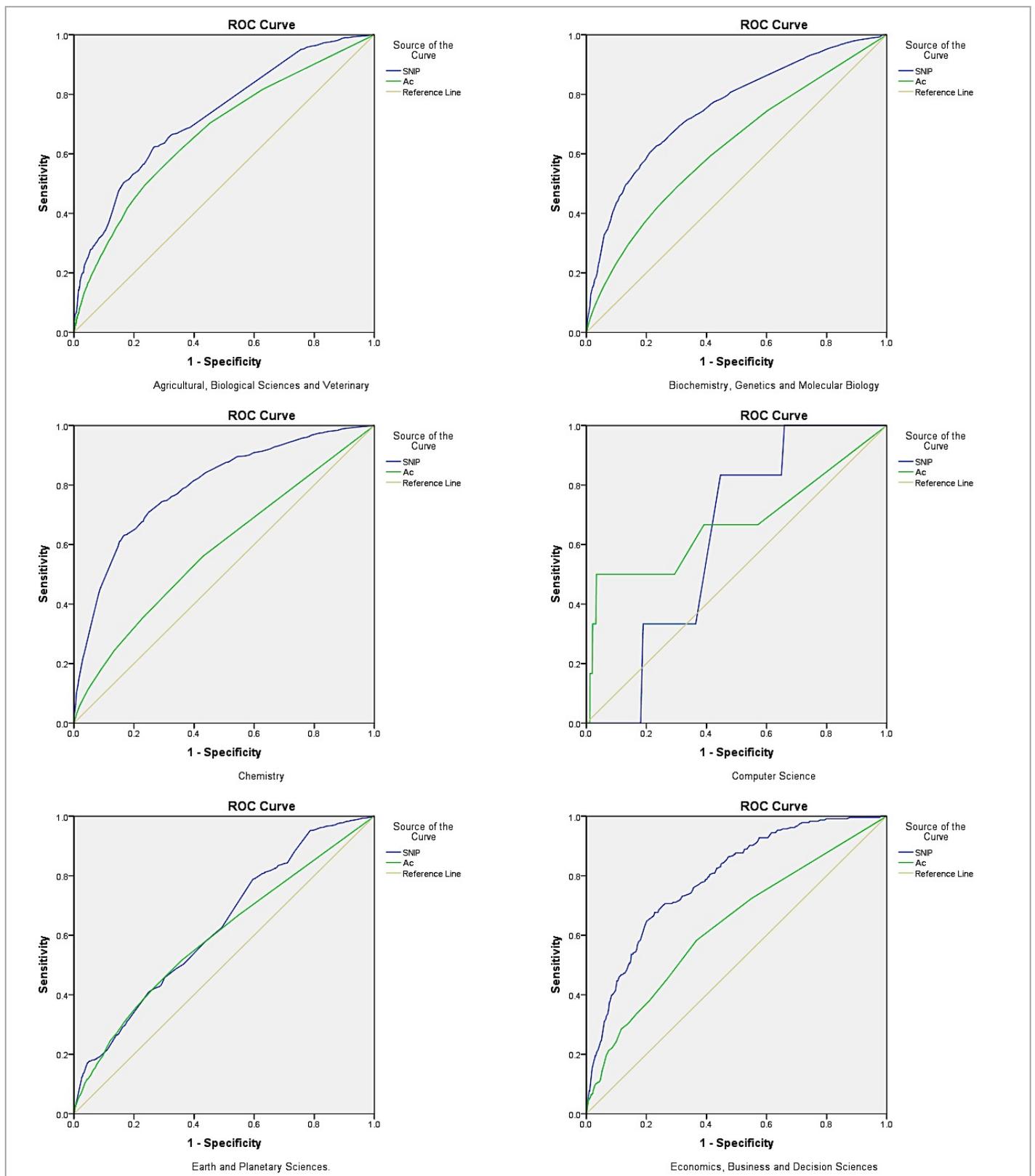

**Figure B-1:** ROC curve of SNIP and Ac to discriminate HC 1% papers Agricultural, Biological Sciences & Veterinary through Economics, Business and Decision Sciences

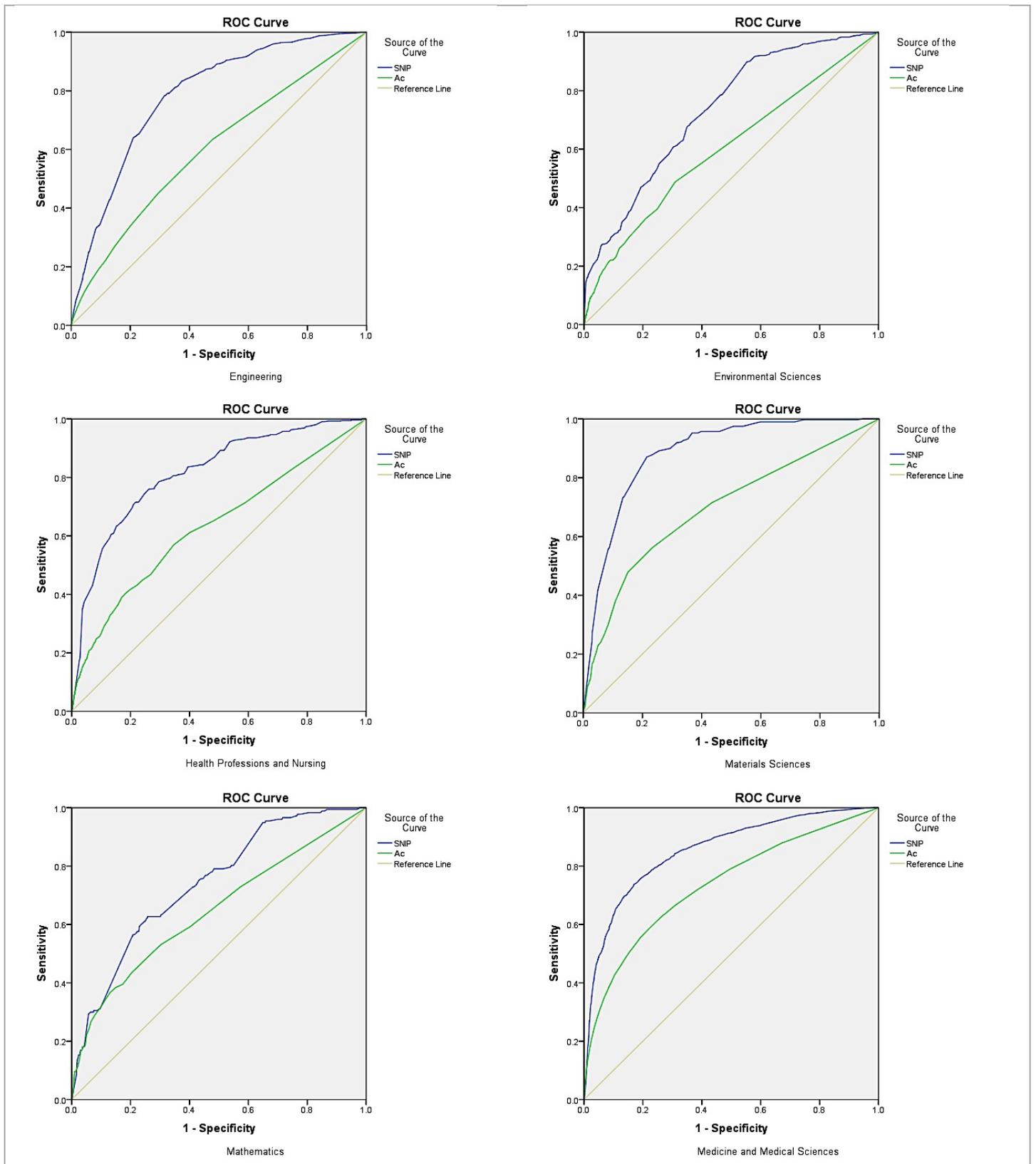

**Figure B-2:** ROC curve of SNIP and Ac to discriminate HC 1% papers Engineering through Medicine and Medical Sciences

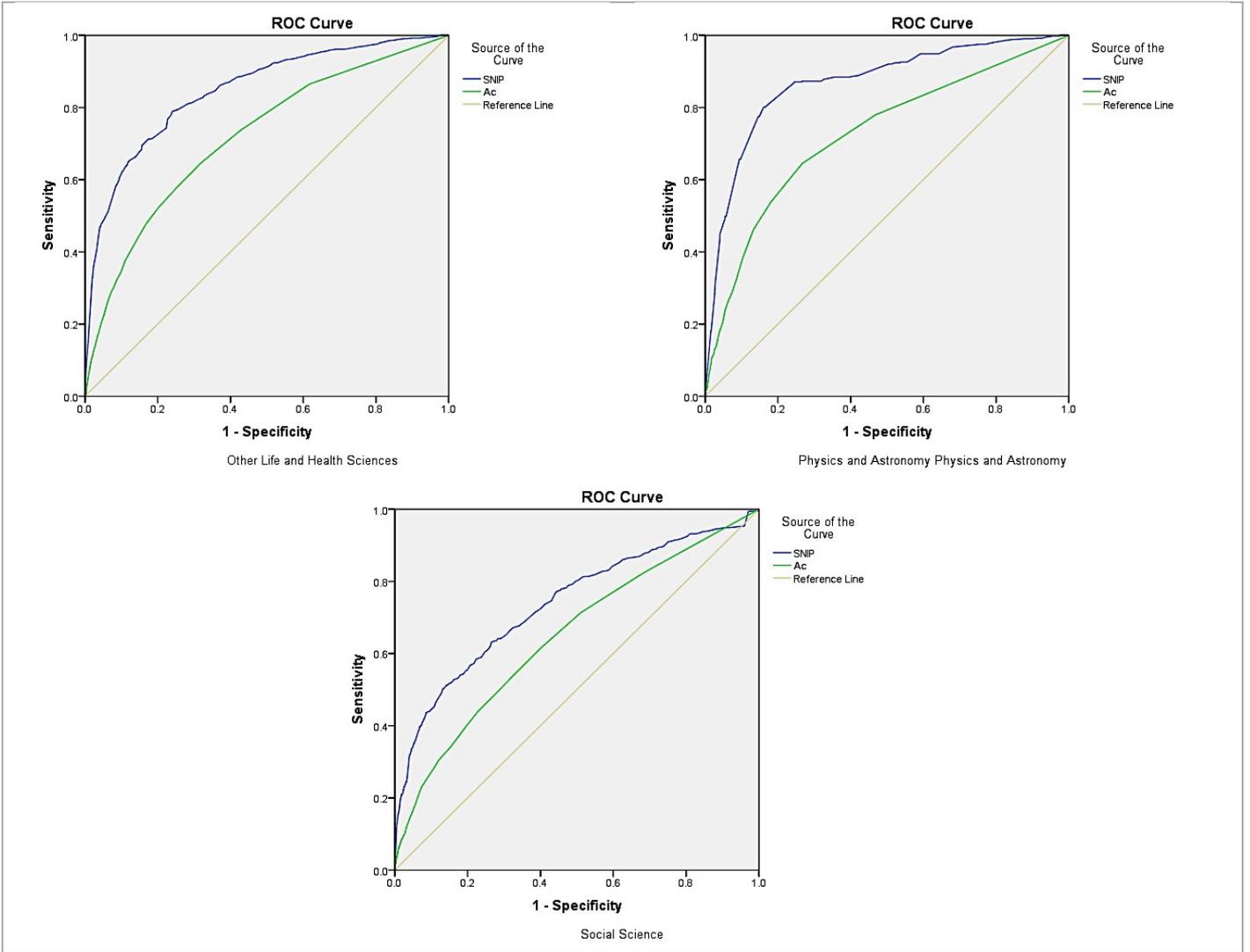

**Figure B-3:** ROC curve of SNIP and Ac to discriminate HC 1% papers Other Life and Health Sciences through Social Sciences